\documentclass[twoside]{article}
\usepackage{fbc,ebc}
\usepackage{cite}
\usepackage{epsfig}

\def\z{${\rm Z}^0$}
\def\ep{e$^+$e$^-$}
\def\ezh{{{\ep} $\to$ {\z} $\to {hadrons}$}}
\def\ezhc{{{\ep}$\to${\z} $\to{hadrons}$}}
\def\eh{{{\ep} $\to {hadrons}$}}
\def\ie{{i.e.}}
\def\eg{{e.g.}}
\def\ea{{et al.}}
\def\col{Collab.}
\def\JT{{\sc Jetset}}
\def\HW{{\sc Herwig}}
\def\OP{OPAL}
\def\DE{DELPHI}

\def\al{\langle}
\def\ar{\rangle}

\def\vs{\vspace*}
\def\hs{\hspace*}
\def\bea{\begin{eqnarray}}
\def\eea{\end{eqnarray}}
\def\be{\begin{equation}}
\def\ee{\end{equation}}
\def\la{\label}
\def\bga{\left( \begin{array}}
\def\ena{\end{array} \right)}

\def\ct{\cite}
\def\bi{\bibitem}
\def\ni{\noindent}

\def\pT{p_T}
\def\phi{\Phi}

\def\3d{y$$\times$$\phi$$\times$$\pT}
\def\lsim{\:{\stackrel{<}{_\sim}}\:}
\def\jour#1#2#3#4{{#1} {#2} (19#3) #4}

\def\jourm#1#2#3#4{{#1} {#2} (20#3) #4}
\def\jourm#1#2#3#4{{#1} {#2} (20#3) #4}
\def\PRp{Phys. Reports}
\def\PRD{Phys. Rev. {D}}
\def\PRC{Phys. Rev. {C}}
\def\PRL{Phys. Rev. Lett. }
\def\AP{Acta Phys. Pol. {B}}
\def\ZP{Z. Phys.  {C}}
\def\EPJ{Eur. Phys. J. {C}}
\def\IJ{Int. J. Mod. Phys. {A}}
\def\NIM{Nucl. Instr. Meth. {A}}
\def\CP{Comp. Phys. Comm.}

\def\PL{Phys. Lett.  {B}}
\def\NPA{Nucl. Phys.  {A}}
\def\NPB{Nucl. Phys.  {B}}

\def\UFN{Physics-Uspekhi}

\title{The many sources effect on
the genuine multihadron correlations \thanks{Invited 
talk presented by
G. Alexander at the 9th International Workshop on Multiparticle Production,
Turin, Italy, June 12 - 17, 2000.}
}

\author{Gideon Alexander\address[TAU]{School of Physics and Astronomy,
The Raymond and Beverly Sackler Faculty of Exact Sciences,\\
Tel-Aviv University, IL-69978 Tel-Aviv, Israel}%
                        \thanks{Email address: alex@lep1.tau.ac.il}
and 
Edward K.G. Sarkisyan\addressmark[TAU]\thanks{Email
address: edward@lep1.tau.ac.il}}

\begin{document}


\begin{abstract}
 Here we report on a study aimed to explore the dependence of the
genuine multiparticle correlations on the number of sources while the
influence of other possible factors affecting the multihadron production is
avoided. The analysis utilised the normalised 
cumulants, calculated
in three-dimensional phase space, of the reaction {\ezhc} using a 
large Monte Carlo event sample.
The multi-sources reactions were simulated by overlaying a few independent
single {\ep} annihilation events.
It was found that as the number of sources $S$ increases, the cumulants do
not change significantly their structure, but those of an order $q > 2$
(i.e. more than 2 pions)
decrease fast in their magnitude.
This reduction can be understood in terms
of combinatorial considerations of source mixing which dilutes 
the correlations by a factor of about $1/S^{q-1}$ which can also serve as
a method to estimated the number of sources.
This expected suppression is well
reproduced by recent cumulant measurements in hadron and
nucleus induced reactions both in one (rapidity) and two
(rapidity vs. azimuthal angle) dimensions.
The diminishing genuine correlations 
effect should also appear in other dynamical
correlations like the Bose-Einstein in  
e$^+$e$^-\to$W$^+$W$^-$ $\to hadrons$ and in 
nucleus-nucleus reactions.  
 \end{abstract}

\maketitle



\section{INTRODUCTION}

 During the last decade an increase interest has been shown for
the genuine multiparticle correlations in multihadron final states of
hadronic, {\ep} and other reactions \ct{revi}.
 Recently OPAL, in its study of {\ep} annihilations on the Z$^0$ mass, has
established the existence of strong genuine multihadron correlations up to
the fifth order \ct{Ogc}.
 In hadron-hadron, like proton-proton, collisions the correlations of more
than three particles have also been observed \ct{hAc,na22,hhc}.
 In contrast to this situation, in heavy ion collisions, at low energies
and/or in reactions of light nuclei, genuine correlations are found to
have non-zero values only up to the third order \ct{AAc2}. 
 Furthermore it has been found out that in general these correlations
become weaker as the reaction average multiplicity increases.
 In nucleus-nucleus collisions at high energies, of tens and hundreds
GeV per nucleon, the two-particle correlations are the only one that
survive \ct{hAc,nfcd,isn,AAc1}.

 This correlation dependence on the average multiplicity is very similar
to the one observed in the investigations of multiparticle dynamical
fluctuations, i.e. variation of many particle bunches in restricted phase
space regions \ct{revi}.
 In these studies, known as intermittency analyses, the observed average
multiplicity dependence of the correlations has been 
proposed to be the consequence of a
mixing of several independent emission sources \ct{is1,is2,isn}.
 As a result, the dynamical fluctuations in nucleus-nucleus collisions are
already well accounted for by two-particle correlations \ct{nfcd,ncf},
whereas in hadron-hadron interactions \ct{hhcf,na22} and in {\ep}
annihilations \ct{Ogc} higher order genuine correlations do exist.

 An analogous situation seems also to exist in the
Bose-Einstein correlations
(BEC) where identical bosons are correlated when they emerge from
the interaction in nearby phase space.
 A genuine three-pion BEC has been detected in hadron-hadron reactions
\ct{hhbec} and found to be even more pronounced in {\ep} annihilations
\ct{eebec}.
 On the other hand the NA44 collaboration \ct{AAbec} has not found, 
in their study of sulphur-lead collisions at 200 GeV/A, 
any genuine three-pion BEC 
as the three-body correlations were 
well reproduced in terms of two-particle BEC. In a recent
reprint, the WA98 collaboration \ct{wacollab} reported on their study of
central lead-lead collisions at 158 GeV/c per nucleon where an evidence
for a three-pion BEC seems to exist. However it was pointed out that,
at least some of the BEC analyses carried out with the 
super proton synchrotron at CERN,
may have to be reevaluated \ct{borghini}.
 Since the intermittency phenomenon and BEC seem to be closely related
\ct{revi}, the dependence of many sources on the strength of the BEC
cannot be excluded.
 The superposition of emitters may also be a reason for the suppression of
BEC of hadrons produced from W-boson pairs in {\ep} annihilations at LEP2
energies where the overlapping of hadrons affect the accuracy of the W
mass measurements \ct{WWbec}.

 All this, as well as the obvious intrinsic interest in the genuine
correlations which carry most of the dynamics of the hadron production
process, points to the need of dedicated studies aimed to investigate the
correlation dependence on the number of emission sources.
 Here we study this dependence by grouping several {\eh}
events to represent a multi-emission sources of particles.
 To obtain significant results, even when only few sources are
considered, one needs a very high statistics, like that which can be
supplied by a Monte Carlo (MC) generated sample. This allows to minimise 
the calculation error and thus be sensitive to the 
correlations dependence on the number of sources.
 Another advantage of using MC generated events is in its possibility 
to generate a multihadron sample free from 
contamination of other processes like, for
example, {\ep} $\to \tau^+\tau^- \to hadrons$.

 Here we present the results of a MC study on the effect of
several emission sources on the genuine higher order multiparticle
correlations \cite{as_plb}.
 The study was based on a generated sample of about 
$5 \times 10^6$ events of the
reaction {\ezh} which passed a full simulation of the OPAL detector at
LEP and did reproduce 
rather well the measured genuine high order correlations present in the
OPAL hadronic {\z} decay data \ct{Ogc}.
Moreover, the {\ezh} annihilations should    
represent well the  
one emission source situation in contrast to events produced in
hadron-hadron interactions. 
Our analysis on the dependence of many sources
on the genuine correlations was thus carried out in a way that avoided  
effects coming from other features, like the multiplicity 
which was discussed recently in connection with 
two-particle BEC analyses in \ct{bec-is} and \ct{bec-iss}.
Inasmuch that final state interactions
between hadrons coming from different sources
can be neglected,
our correlation study based on {\ezh} annihilations 
may be extended to other types of reactions since the
hadronisation process is believed to rest on a common basis \ct{qcd}.

\section{THE ANALYSIS METHOD}

 The analysis is based on a generated sample of hadronic Z$^0$ decays
using the {\JT} 7.4 MC program \ct{JT} including a full simulation of the
OPAL detector at LEP \ct{MCOd}.
 The MC sample also included initial-state radiation and effects of finite
lifetimes.
 The parameters of the program were tuned to yield a good description of
the measured event shape and single particle distributions \ct{MCO}.

 The selection criteria for multihadron events used here are identical to
the ones previously utilised by OPAL in their recent data analysis of
multiparticle correlations \ct{Ogc}.
 In particular, selected events were required to have at least five
charged tracks each having at least 20 measured
points in the jet chamber where the first point had
to be closer than 40 cm from the beam axis.
 The cosine of the polar angle of the event sphericity axis with respect
to the beam direction was required to be less than 0.7 to ensure 
that the event
lies within the volume of the detector.
 The sphericity axis was calculated by using all accepted tracks and
electromagnetic and hadronic calorimeter clusters.

 To simulate several emission sources we did overlay several {\ezh}
generated events and analysed the correlations between pions as if they
were created in a single event.
 The kinematic variables are defined within each generated {\ep} event
with respect to its own sphericity axis.
 For correlation analyses of variables like rapidity this procedure is
equivalent to the one where the events are rotate to a common sphericity
axis.
 This simulates multi-sources' events producing hadrons, here taken to be
pions, emerging from a common emitter. To note is that
in this procedure 
the average event multiplicity is directly proportional
to the number of sources. In our analysis each generated event was
used only once to avoid correlations between different sources, 
which for this reason required a very large MC event sample.

 To extract the genuine dynamical $q$-particle correlations, we used
bin-averaged normalised factorial cumulant moments, or cumulants, first
proposed in Ref. \ct{cum} as a tool for 
the search of genuine multiparticle correlation,

\be
K_q={1\over M}
\sum_{m=1}^{M} \int_{\delta y} \prod_i {\rm d}y_i 
\frac 
 {C_q(y_1,\ldots, y_q)}
 {[ \int_{\delta y} {\rm d}y\rho_1(y)] ^q}\:.
\la{kmy}
\ee
\ni 
The $C_q(y_1,\ldots, y_q)$ are the $q$-particle correlation functions
given by the inclusive $q$-particle density distributions
$\rho_q(y_1,\ldots, y_q)$ in terms of cluster expansion, {\eg},

$$
{\hs{-5.4cm} C_3(y_1,y_2,y_3)}
$$
$$
{\hs{-.5cm}
=\rho_3(y_1,y_2,y_3)
    -\sum_{(3)}\rho_1(y_1)\rho_2(y_2,y_3)}
$$
\be
\qquad 
  +\,2\,\rho_1(y_1)\rho_1(y_2)\rho_1(y_3)\:.
\la{cr}
\ee

\ni
 Here $M$ is the number of equal bins, having a width $\delta y$, 
into which
the event phase-space is divided and the subscript (3) denotes the
number of permutations.
 For simplicity we show all formulae in 1-dimensional ({\eg}, rapidity)
phase space.

 The feature of the $C_q$-functions is that they vanish whenever there are
no genuine correlations, {\ie}, the correlations are due to those present
in lower orders.
 The correlations extracted are of a dynamical nature since the cumulants
share with normalised factorial moments (the intermittency analysis tool)
the property of statistical noise suppression.

Here we computed the cumulants as they are used in experimental
studies, in particular we used the form applied in Ref. \ct{Ogc} namely,

\be 
K_q= 
\frac{
{\cal N}^q \cdot 
\sum_{m=1}^{M} k_q^{(m)}
}{ 
\sum_{m=1}^{M} N_m(N_m-1)\cdots(N_m-q+1)
}\:\;. 
\la{kmh}
\ee 
 The $k_q^{(m)}$ factors are the unnormalised factorial cumulant
moments, or the Mueller moments \ct{math}, calculated for the $m$th bin.
 These factors represent the correlation functions $C_q$ integrated over
the bin and $N_m$ is the number of particles in the $m$th bin summed over
all the $\cal N$ events.
 The definition (\ref{kmh}) takes into account the non-uniform shape of
the single-particle distribution and the bias when the cumulants are
computed at small bins.

 The cumulant calculations were performed in the three-dimensional phase
space of the kinematic variables commonly utilised in this kind of studies
\ct{revi}, namely:

 \begin{itemize}
 \item The rapidity, $y=\ln \sqrt {(E+p_{\|})/(E-p_{\|}) }$, with $E$ and
$p_{\|}$ being the energy and longitudinal momentum of the hadron within
the interval $-2.0 \leq y\leq 2.0$;
 \item The transverse momentum in the interval $0.09 \leq \pT\leq 2.0$
GeV/$c$;
 \item The azimuthal angle, $0\leq\phi<2\pi$, calculated with respect to
the eigenvector of the momentum tensor having the smallest eigenvalue, in
the plane perpendicular to the sphericity axis.
 \end{itemize} 
 These variables are defined with respect to the sphericity axis, in a way
and within the intervals similar to those used in a recent OPAL analysis
\ct{Ogc} and in other cumulant studies \ct{revi}.

\section{GENUINE CORRELATIONS AND THE NUMBER OF SOURCES}

\begin{figure*}[!htb]
\vs{6.cm}
\hs{.9cm}
\epsfysize=8.7cm
\epsffile[20 150 200 500]{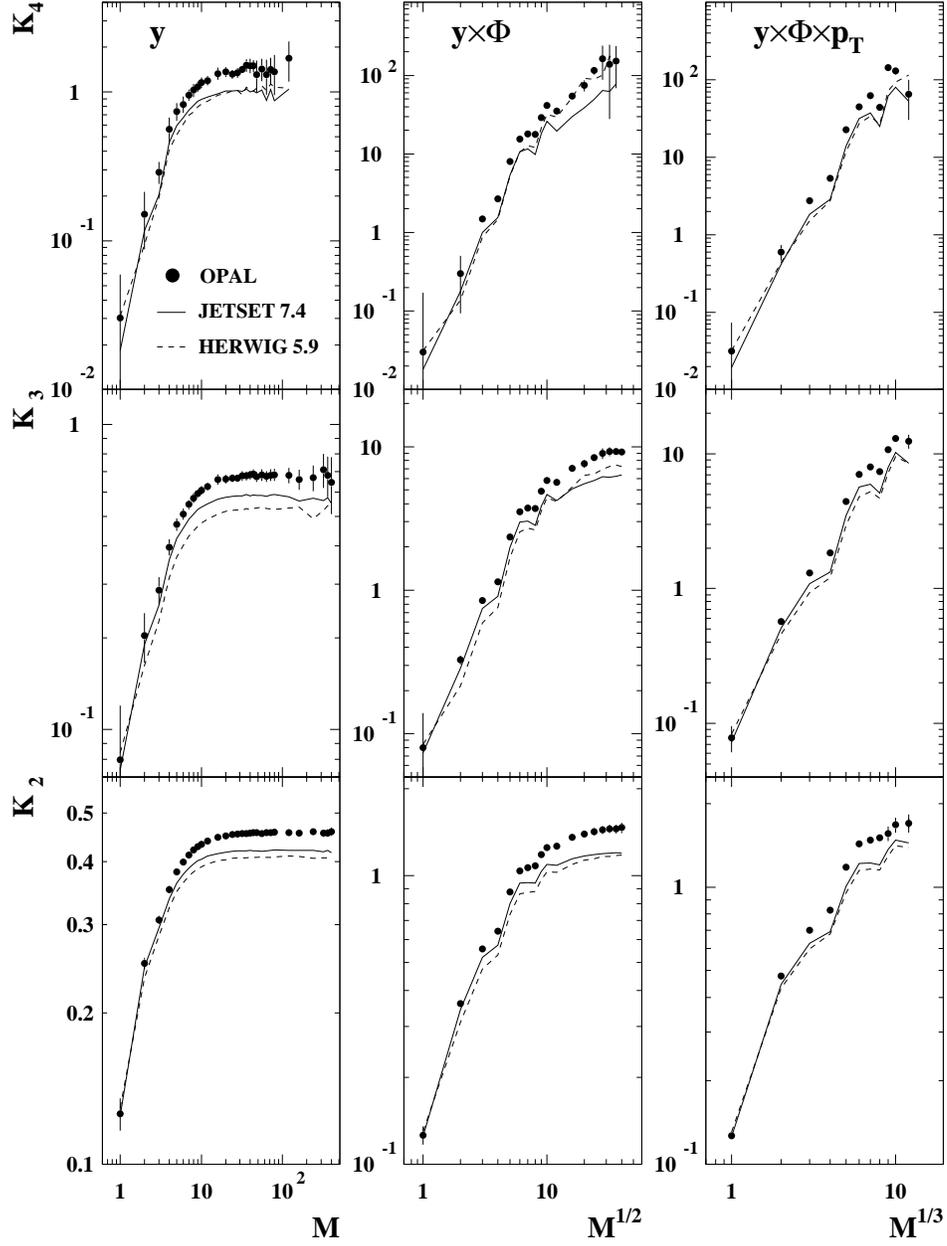}
\vs{1.6cm}
\caption{ 
 Cumulants of order $q =$ 2, 3 and 4 as a function of $M^{1/D}$, where $M$
is the number of bins of the $D$-dimensional sub-spaces of the phase space
of rapidity ($y$), azimuthal angle ($\phi$) and transverse momentum
($\pT$), compared to two MC models.
The data and the MC predictions are taken from Ref. \ct{Ogc}.
}
\la{cm24}
\end{figure*}

\subsection{Monte Carlo studies}

\begin{figure*}[!htb]
\vs{4cm}
\epsfysize=10.7cm
\epsffile[20 150 200 500]{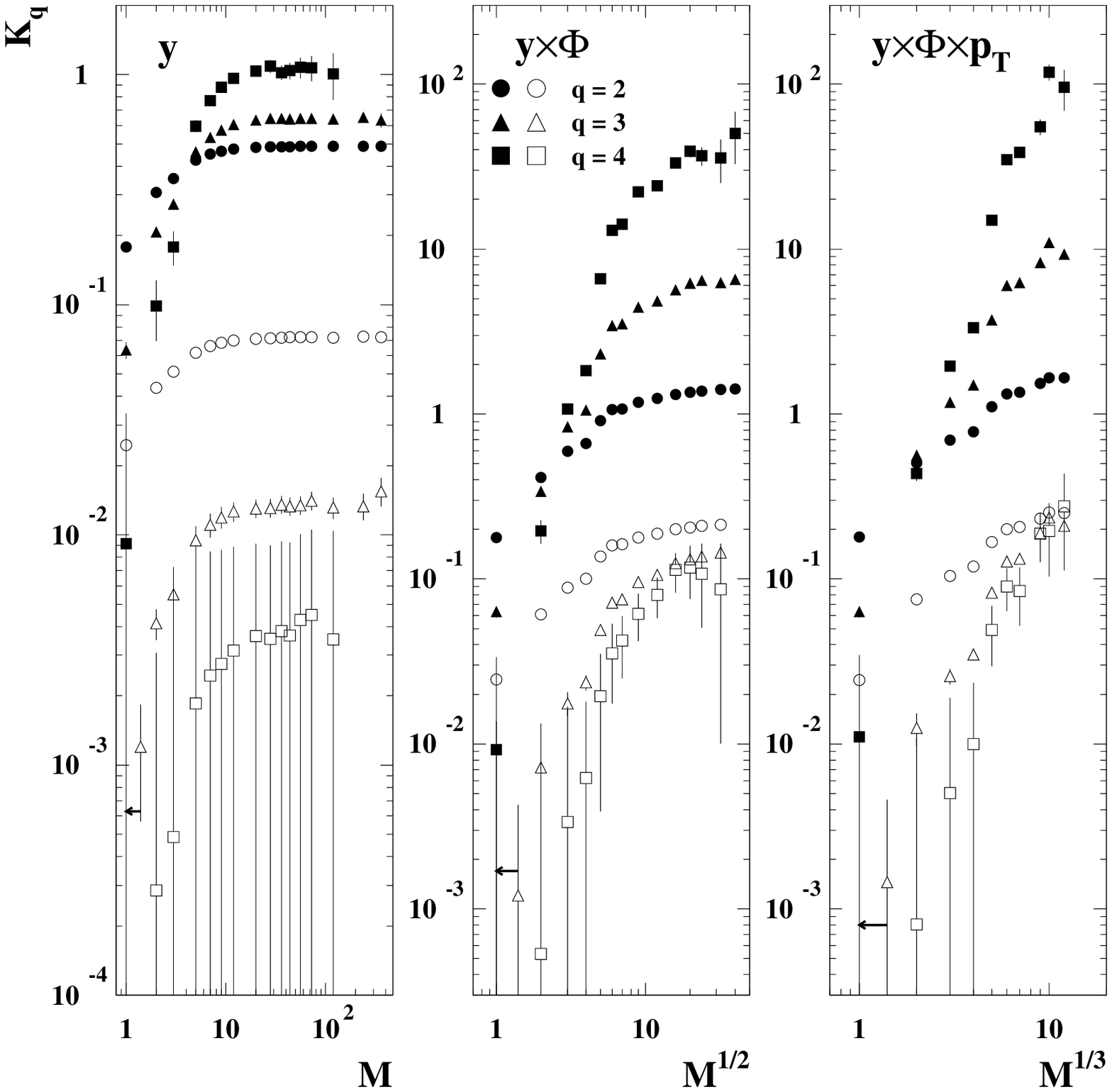}
\caption{\
The MC predicted cumulants of order $q =$ 2, 3 and 4 as a function of
$M^{1/D}$, where $M$ is the number of bins of the $D$-dimensional
sub-spaces of the phase space of rapidity ($y$), azimuthal angle ($\phi$),
and transverse momentum ($\pT$).
 The solid symbols represent the cumulants for a single source, while the
open symbols are the cumulants values of seven sources.
 }
\la{cm24s}
\end{figure*}

\smallskip
 In Fig. \ref{cm24} we reproduce the cumulants of orders $q=2,$ 3 and 4
measured by {\OP} and compared with those based on two MC models,
{\JT} 7.4 and {\HW} 5.9 \ct{Ogc}.
 The cumulants were calculated in 1-dimensional sub-space of rapidity, in
2-dimensional rapidity vs. azimuthal angle sub-space and in
3-dimensional phase space of rapidity, azimuthal angle and transverse
momentum.
 The measured cumulants are seen to have large non-zero values thus
inferring strong genuine multiparticle correlations down to very
small bin sizes.
 As was also concluded by OPAL \ct{Ogc}, the high order
fluctuations cannot be reproduced by lower-order correlations and indeed
require also high-order correlations to be present.
 Thus higher order correlations do play an important role in the
hadronic Z$^0$ decays processes.

As can be seen from Fig. \ref{cm24}, although the MC cumulants slightly
underestimate the data starting at intermediate bin size, they 
reproduce well the over all behaviour of the cumulants as a 
function of the bin-size.
This fact is utilised here for the study 
of the genuine multiparticle correlations dependence on the number of sources.

\begin{figure}[!thb] 
\vs{4.2cm} 
\hs{.5cm} 
\epsfysize=9.6cm
\epsffile[45 150 200 500]{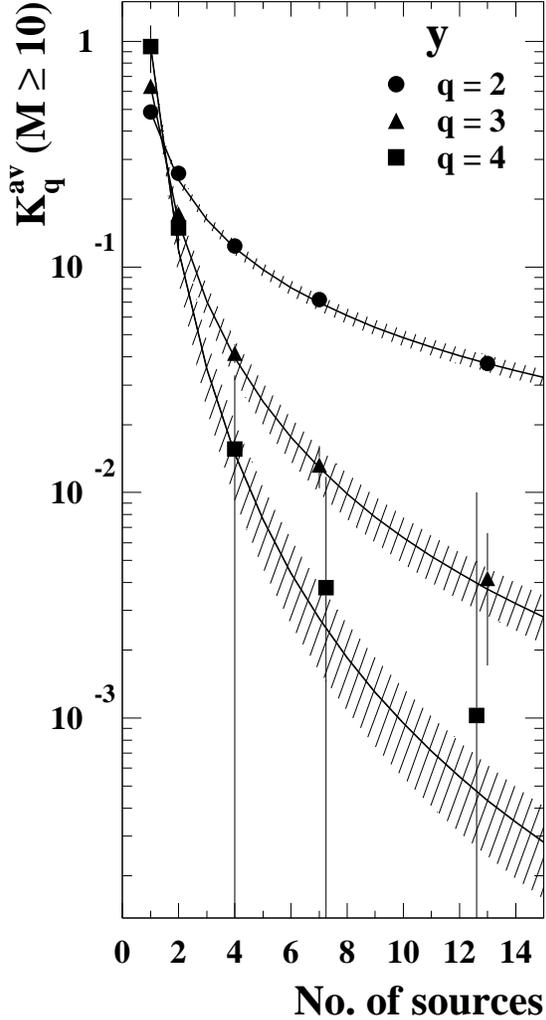} 
\caption{
 The dependence of the averaged 1-dimensional rapidity cumulants $K_q^{\rm
av}$ of order $q=$ 2, 3 and 4 on the number of {\ep} sources.
 The cumulants were averaged over the $M$-range where they 
maintain an almost constant value (see Fig. 1).
 The lines represent the expected dilution according to Eq. (\ref{difi})
where $q$ is neglected in comparison to the multiplicity $n$.
 The striped areas are the allowed regions when $q$ is not neglected with
respect to $n$ (see text).
 }
\la{1c}
\end{figure}

 In Fig. \ref{cm24s} we compare the MC based cumulants of orders $q=2,$ 3
and 4 calculated from a single {\ezh} events  (solid symbols) with those
obtained by overlaying seven such events to represent seven hadronic
sources (open symbols).
 The calculations were performed in the 3-dimensional phase space of
rapidity, azimuthal angle and transverse momentum as well as in its
projections onto 1-dimensional rapidity and onto 2-dimensional rapidity
vs. azimuthal angle sub-spaces.

 The following observations can clearly be made from Fig. \ref{cm24s}. 

 \begin{itemize} 
\item 
 The existence of a dynamical component, {\ie} rise of the cumulants with
increasing number of bins $M$, is seen to be present both in the single
source as well as in the case of many sources.
 Although the slopes of this scaling behaviour are smaller for several
sources than for a single source, they are still strongly present.         
 It is also evident that the scaling character is kept as the
number of sources increases. For example, the single-source 1- and
2-dimensional cumulants level off at the same $M$ values as those for
seven sources.
  No such saturation exists for the one and seven sources cumulants in
three dimensions.

\item 
The genuine dynamical correlations, measured by the cumulants,
significantly decrease as the number of sources increases.
 This decrease is stronger for higher order correlations. Whereas
the two-particle cumulants suffer a reduction of an order of magnitude
as the number of sources increases from one to seven, the four-particle
cumulants diminish by three or four orders of magnitude.

\item 
 The hierarchy of the $K_q$ cumulants is reversed as the number of sources
increases.
  The cumulants derived from the single-source events increase with
increasing $q$-order so that $K_2^{(1)} < K_3^{(1)} < K_4^{(1)}$, whereas
the hierarchy in
the cumulants calculated for seven sources is reversed
namely, $K_2^{(7)} > K_3^{(7)} \geq K_4^{(7)}$.
 In addition, the multi-sources cumulants of order $q>2$ have almost the
same reduced value namely, $K_3^{(7)} \approx K_4^{(7)} \lsim {\cal
O}(0.1)$. 
 This last feature does not change as the dimension increases.

\item 
 The overall dominant feature of the analysis results is the diminishing
value of the higher order cumulants as the sources number increases
leaving the $K_2$ to be  the dominant genuine multiparticle
correlation.
 \end{itemize}

\begin{figure}[!htb] 
\vs{4.2cm} 
\hs{0.5cm} 
\epsfysize=9.6cm
\epsffile[45 150 200 500]{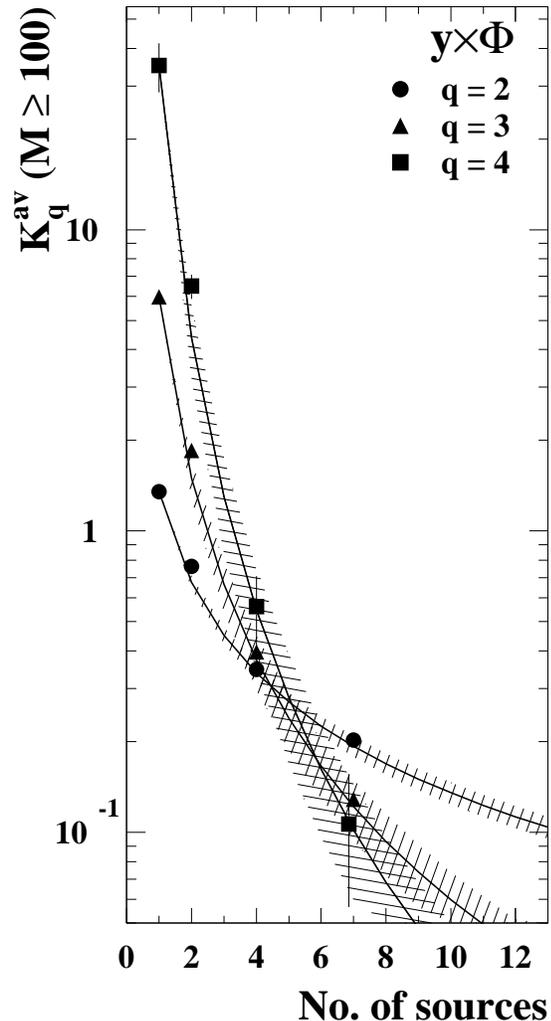} 
\caption{ 
The same as Fig. \ref{1c} but for the
2-dimensional sub-space of rapidity and azimuthal angle.
 }
\la{2c}
\end{figure}

To analyse further the observed diminishing effect, the cumulants
were averaged over the $M$ region where they are
seen in Fig. \ref{cm24s} to reach an almost constant value, {\ie} over
$M\ge10$ in one dimension (rapidity) and $M\ge100$ in two dimensions
(rapidity vs. azimuthal angle).
 The resulted $M$-averaged cumulants $K_q^{\rm av}$ are shown in Figs.
\ref{1c} and \ref{2c} respectively for one and two dimensions.
 The corresponding values are also listed in Table \ref{t1c} for the
1-dimensional cumulants and in Table \ref{t2c} for the 2-dimensional ones
together with the OPAL measured data cumulants \ct{Ogc} of single {\ezh}
events.

 As  Table \ref{t1c} shows, the 1-dimensional data cumulants 
do agree with those derived from the MC sample for $q\leq3$. 
 Even though the single-source MC based $q = 4$ $M$-averaged 
cumulant lies lower
than the measured points it is still consistent within errors
with the data.
 This however is not the case for the two dimensions 
as can be deduced from the values given in Table
\ref{t2c}. Already for $q \geq 3$ the measured and the MC based 
2-dimensional averaged cumulants
disagree. 
 This is due the faster increase of the data cumulants compared to those
obtained from MC (see Fig. \ref{cm24}).

\begin{table*}[htb]
\caption{
 The Monte Carlo $M$ averaged 1-dimensional rapidity $K_q^{\rm av}$
cumulants obtained in the {\ezh} reaction compared to those 
measured by OPAL \ct{Ogc} in single data events.
 The averages were taken over the $M$ region
where the cumulants reached a constant value (see Fig. \ref{cm24s}).
 }
\la{t1c}
\renewcommand{\tabcolsep}{1.8pc} 
\renewcommand{\arraystretch}{1.17} 
\begin{tabular}{ccccc} \hline  
{\small No. of} & \multicolumn{3}{c}{$K_q^{\rm av}\,
   (M\geq10)$}\\\cline{2-4} 
{\small sources} & $q=2$ & $q=3$& $q=4$ & 
   {\raisebox{1.3ex}[0cm][0cm]{Sample}} \\ \hline
\smallskip
 & $0.45\pm 0.01$ & $ 0.67\pm 0.04$ & $1.36\pm 0.21$& OPAL data \\
1 & $0.486\pm 0.002$ & $0.632\pm 0.017$ & $0.950 \pm 0.225$ & MC\\
2 & $0.260\pm 0.001$ & $0.171\pm 0.006$ & $0.147 \pm 0.034$ & ''\\
4 & $0.124\pm 0.001$ & $0.041\pm 0.003$ & $0.016\pm 0.012$ & ''\\
7& $0.072\pm 0.001$ & $0.013\pm 0.003$ & $0.003\pm 0.008$ & ''\\
13\ \ & $0.037\pm 0.001$ & $0.004\pm 0.003$& $0.001\pm0.009$  & ''\\
\hline
\end{tabular}\\[2pt]
\end{table*}

\begin{table*}[htb]
\caption{
 The Monte Carlo $M$ averaged 2-dimensional (rapidity vs. azimuthal
angle) $K_q^{\rm av}$ cumulants obtained in the {\ezh} reaction compared
with those obtained in a recent OPAL measurements \cite{Ogc} 
of single data events.
 The averages were taken over the $M$ region
where the cumulants reached an almost constant value (see Fig. \ref{cm24s}).
 }
\la{t2c}
\renewcommand{\tabcolsep}{1.9pc} 
\renewcommand{\arraystretch}{1.19} 
\begin{tabular}{ccccc} \hline  
{\small No. of} & \multicolumn{3}{c}{$K_q^{\rm av}\,
   (M\geq100)$}\\\cline{2-4} 
{\small sources} & $q=2$ & $q=3$& $q=4$ & 
   {\raisebox{1.3ex}[0cm][0cm]{Sample}} \\ \hline
\smallskip
 & $1.39\pm 0.04$ & $ 7.91\pm 0.36$ & $100\pm 52$& OPAL data\\
1 & $1.353\pm 0.005$ & $6.0\pm 0.2$ & $35.0 \pm 6.5$ & MC\\
2 & $0.763\pm 0.002$ & $1.85\pm 0.03$ & $6.5 \pm 0.6$ & ''\\
4 & $0.347\pm 0.001$ & $0.398\pm 0.012$ & $0.6\pm 0.2$ & ''\\
7& $0.202\pm 0.001$ & $0.129\pm 0.006$ & $0.11\pm 0.05$ & ''\\
\hline
\end{tabular}\\[2pt]
\end{table*}

 From the features of the $M$-averaged cumulants shown in
Figs. \ref{1c} and \ref{2c} one can conclude with the following obvious
points.

 \begin{itemize}
 \item
 In one, as well as in two dimensions, the 
cumulants of order $q>2$ decrease fast with the increased number of
sources.
 \item 
 In one dimension the hierarchy is reversed already for two sources and the
two-particle correlations visibly dominate over the higher order ones.
 \item 
 In two dimensions the hierarchy changes too, but at 
a higher number ($>$ 2) of sources 
since additional correlations in the second variable
(in our case the azimuthal angle) play also a  role. 
The latter correlations are
within a jet which are suppressed when all the phase
space is projected into one direction, rapidity or azimuthal angle, as 
is discussed in \ct{Ogc}.
Thus it seems that there are features in the second dimension, here
the azimuthal angle, that lead to higher order , $q > 2$, correlations
which do not contribute to the 1-dimensional
rapidity correlations.  
 \item
 At higher number of sources the dominant role of the two-particle
correlations becomes more pronounced.
 \end{itemize}
\smallskip

\subsection{Correlation dilution due to source mixing}

\smallskip
 Within the procedure adopted here for the simulation of multi-source
events, it is clear that if a genuine correlation exists it can only be
detected in groups of $q$ pions emerging from the very same source.
 In those $q$-group combinations which emerge from at least two sources,
genuine correlations should not be present.
 This means that for $K_q$ cumulants that are calculated over all possible
$q$-pion groups, the higher the number of sources the more diluted will
be the signal for genuine correlations.

For a given $q$-order the genuine correlation dilution factor is thus:

\be
R_q\ =\ \frac{P_q^{\rm G}}{(P_q^{\rm G} +P_q^{\rm NG})}\ ,
\la{dif}
\ee        
 where $P_q^{\rm G}$ denotes the number of $q$-particle groups, e.g.,
pairs or triplets of pions, which emerge from the same source.
 The term $P_q^{\rm NG}$ stands for the number of all possible
combinations of $q$-particle groups which emerge from at least two
sources.

 Since all sources are produced in the same reaction and at the same
energy, they do have an identical average charged multiplicity.
 For the estimation of $R_q$ we assume that all the $S$ sources have 
the same fixed charged multiplicity $n$. 
 In this case one has $P_q^{\rm G} = S\, {n \choose q}$, and the dilution
factors at $q=2$, 3 and 4 are given by

\bea
\nonumber
R_2& = & \frac{ {n \choose 2}\,S}
                {{n \choose 2}\, S+n^2\,{S \choose2}
              }\ \
{\stackrel{n\gg 1}{\longrightarrow}}\ \ \frac{1}{S}  \:\, ,
\\ \nonumber 
\\ \nonumber
R_3& = & \frac{ {n \choose 3}\,S}
    {{n \choose 3}\,S+n^3\, {S\choose 3}+2\,n\,{n\choose 2}{S\choose 2}
                } \ \
{\stackrel{n\gg 2}{\longrightarrow}}\  \ \frac{1}{S\,^2} \:\,,
\\ \nonumber
\\ \nonumber
R_4& = & 
\nonumber 
\eea 
$$
\frac{ {n \choose 4}\,S}
                {{n \choose 4}\,S +
     n^4\,{S\choose 4}+3\,n^2\,{n\choose 2}{S\choose 3} +
     {n\choose 2}^2\ {S\choose 2}+ 2\,n\,{n\choose 3}{S\choose 2}}
$$
\be
\qquad 
{\stackrel{n\gg 3}{\longrightarrow}}\ \ \frac{1}{S\,^3} \:\,,
\la{difn} 
\ee
 where the denominators include the number of all possible $q$-particle
combinations in $S$ sources of charged multiplicity $n$.
  These dilution factors dependence on the number of sources can also be
derived in terms of cumulants \ct{is1}. 

 From these $R_q$ relationes follows that as long as $n \gg q$ one
has a general expression for the dilution factor namely,

\be
R_q\  \ {\stackrel{n\gg q}{\longrightarrow}}\ \ 
            \frac{1}{S\,^{q-1}}\:.
\la{difi}
\ee

 To compare the dilution factors $R_q$ with our correlation results shown
in Figs. \ref{1c} and \ref{2c}, they do have to be multiplied by
$K_q^{{\rm av}(1)}$ which is a measure of the genuine $q$-order
correlation present in a single sources.
 The solid lines shown in Figs. \ref{1c} and \ref{2c} thus represent the
diluted cumulants $K_q^{\rm av}=K_q^{{\rm av}(1)}\times R_q$. 
 The striped areas in which the lines are embedded are the allowed regions
when $q$ is not neglected with respect to the multiplicity $n$.

 The agreement between the cumulant calculations and the dilution factors
predictions is really remarkable. 
  In the one-dimensional case the predictions follow the MC based
cumulants for $q=2$ and 3 and are certainly well within the rather large
errors of the $q = 4$ cumulants.
 In two dimensions, in addition to the remarkably good agreement with the MC
cumulants, the dilution factor predictions follow closely also 
the hierarchy change of the $K_q$ moments.

 For the order $q\ =\ 2$ one can relax the fixed charged multiplicity
assumption and allow them to be different and still retain
the $R_q \simeq 1/S^{q-1}$ relation as long as the multiplicity
distribution is of a Poisson nature. This however is not the case
for orders higher than 2. Nevertheless for order 
$q\ =\ 3$ the relation $R_3 = 1/S^2$, derived from the fixed
multiplicity assumption, is still valid as it describes
well the      
$K^{\rm av}_3$ values up to at least thirteen sources (see Fig. \ref{1c}).
The large cumulants' errors associated with the $q\ =\ 4$ order
prohibits to judge how accurate is the $R_4\ =\ 1/S^3$ relation.

 An additional interesting and useful application of the relation $R_q
\simeq 1/S^{q-1}$ is that it offer a method to estimate the average
number
of sources $\al S\ar$ via the cumulant averaged values over the large $M$
region of two sequential $q$-orders through the ratio,

\be
\al S\ar\ \simeq \ 
\frac{K_{q+1}^{{\rm av(1)}}}{K_q^{{\rm av(1)}}}      
\times 
\frac{K_q^{{\rm av}}}{K_{q+1}^{{\rm av}}} \ .
\la{useful}
\ee   
\smallskip

\subsection{Comparison with hadron and nucleus induced reactions}

\begin{table*}[htb]
\caption{
 The $M$ averaged 1-dimensional rapidity cumulants $K_q^{\rm av}\:$ of
orders $q=2$ and 3 measured in several hadronic reactions.
 The cumulants were averaged over the $M$ regions where they were seen to
approach a constant value.
  The quoted values were estimated from the relevant published figures
in the papers listed in the table references.
 }
\la{tw1c}
\renewcommand{\tabcolsep}{1.33pc} 
\renewcommand{\arraystretch}{.93} 
\begin{tabular}{cccllll} 
\hline 
{\raisebox{-1.3ex}[0cm][0cm]{$\al n_{\rm ch}\ar $}}
   & \multicolumn{2}{c}{$K_q^{\rm av}$} &
   {\raisebox{-1.3ex}[0cm][0cm]{Reaction}} & Beam energy & 
   {\raisebox{-1.3ex}[0cm][0cm]{Ref.}} \\\cline{2-3}
   & $q=2$ & $q=3$ &     & (GeV) & \\ \hline
 $\sim 8\ \ \ \ \ \ \ \ \ $ & $0.32\pm 0.02$ & $0.26\pm 0.12$ & $\pi$p &
250 & \ct{hAc}$^*$\\
\vs{.5mm}
21.1 \ \  & $0.21\pm 0.04$  &  $0.14\pm 0.18$  & pEm     & 200 &
   \ct{hAc}$^*$\\ 
\vs{.5mm}
$>50\ \ \ \ \ \ \ $ & $0.34\pm 0.02$ & $0.12\pm 0.03$ &
AuEm & 10.6$A$ & \ct{AAc2}\\ 
\vs{.5mm}
73.3 \ \ & $0.21\pm 0.03$ & $0.05\pm 0.07$ & SiEm & 14.5$A$ &
   \ct{AAc1,AAc2}\\
\vs{.5mm}
81.1 \ \  & $0.20\pm 0.05$  &  $0.00\pm 0.12$  & OEm    & 60$A$ &
\ct{AAc1}\\
\vs{.5mm}
154.9 & $0.11\pm 0.05$  &  $0.01\pm 0.18$  & OEm     & 200$A$ &
  \ct{hAc}$^*$\\
\vs{.5mm}
216.1 & $0.25\pm 0.05$  &  $0.08\pm 0.10$  & SEm    & 200$A$ &
  \ct{AAc1}\\
\vs{.3mm}
272.6 & $0.09\pm 0.05$  &  $0.00\pm 0.18$  & SEm    & 200$A$ &
  \ct{hAc}$^*$\\
\vs{.5mm}
289.8 & $0.11\pm 0.08$  &  $0.02\pm 0.10$  & SAu & 200$A$ &
  \ct{isn}$^{**}$\\
\vs{0.5mm}
355.0 & $0.08\pm 0.05$  &  $0.00\pm 0.08$  & SAu  & 200$A$ &
  \ct{hAc}$^*$\\
\vs{.2mm}
383.9 & $0.07\pm 0.04$  &  $0.00\pm 0.05$  & SAu  & 200$A$ &
 \ct{isn}$^{***}$\\
\hline
\end{tabular}\\[2pt]
{ 
$^*$ Calculations are based on the measured factorial moments.\ \ 
$^{**}$ Semi-central collisions.\ \ 
$^{***}$ Central collisions.}
\end{table*}

\begin{table*}[htb]
\caption{
 The $M$ averaged 2-dimensional (rapidity vs. azimuthal angle) cumulants
$K_q^{\rm av}\:$ of orders $q=2$, 3 and 4 measured in nucleus-nucleus
collisions.
 The cumulants were averaged over the $M$ regions where they were seen to
approach a constant value.
  These quoted values were estimated from the relevant published figures
given in the references listed in the table.
 }
\la{tw2c}
\renewcommand{\tabcolsep}{.98pc} 
\renewcommand{\arraystretch}{.48} 
\begin{tabular}{ccccllll} 
\hline 
{\raisebox{-1.3ex}[0cm][0cm]{$\al n_{\rm ch}\ar $}}
   & \multicolumn{3}{c}{$K_q^{\rm av}$} &
   {\raisebox{-1.3ex}[0cm][0cm]{Reaction}} & Beam energy & 
   {\raisebox{-1.3ex}[0cm][0cm]{Ref.}} \\\cline{2-4} 
\vs{.2mm}
   & $q=2$ & $q=3$ &   $q=4$ &  & (GeV) & \\ \hline 
\vs{.5mm}
  $>50\ \ \ \ \ \ \ $ & $0.79\pm 0.02$  &  $1.3\pm 0.1$ &  $3.2\pm 2.8$ &
AuEm & 10.6$A$ &  \ct{AAc2}\\
\vs{.5mm}
  73.3\ \ \ & $0.20\pm 0.05$ & $0.09\pm 0.03$ & $0.4 \pm 0.3$ & SiEm &
14.5$A$ & \ct{AAc2}\\
\vs{.5mm}
 119.3 & $0.57\pm 0.03$ & $0.3\pm 0.1$ & $0.4 \pm 0.7$ & OEm & 200$A$ &
\ct{AAc1}\\
\vs{.2mm}
 216.1 & $0.25\pm 0.03$ & $0.07\pm 0.04$ & $0.1\pm 0.2$ & SEm & 200$A$ &
\ct{AAc1}\\
 \hline
\end{tabular}
\end{table*}

\smallskip
 As is already mentioned in the introduction, the genuine correlations
measured in {\ep} annihilations \ct{Ogc} are found to be weaker in
hadronic interactions \ct{hhcf,hAc,na22,hhc} and even more so in nuclear
collisions \ct{hAc,nfcd,isn,AAc1,AAc2}.
 In nucleus-nucleus collisions at ultra-relativistic energies only the
second-order correlations are found to have non-zero values in rapidity
\ct{hAc,nfcd,isn,AAc1}, while in two dimensions (rapidity and azimuthal
angle) the third-order cumulants are also detected \ct{AAc1,AAc2}.

 In Table \ref{tw1c} we list the results obtained by several experiments
\ct{hAc,AAc1,isn,AAc2} on the $M$-averaged rapidity cumulant values for
$q=2$ and 3.
 In Table \ref{tw2c} the analogous cumulants for $q=2,$ 3 and 4 are given 
as measured in two dimensions, rapidity vs. azimuthal angle
\ct{AAc1,AAc2}.
 These average values were taken over the $M$-region where the cumulants
are seen in the published figures to reach a constant level.
 The reactions and their cumulants values are ordered according to their
reported mean charged multiplicity, from the lowest value to the highest
one.

 Tables \ref{tw1c} and \ref{tw2c} show that in hadron including nucleus
induced reactions the two-particle correlations decrease rather fast as
the mean multiplicity increases both in one and two dimensions.
 However the three-particle correlations are found to be essentially
non-existing in one dimension even at moderately small mean multiplicity,
while in two dimensions these correlations are seen still to be non-zero. 
 Moreover, the higher the order of the 2-dimensional correlation the
larger the value is. 
 These measurements, particularly for 2-dimensional cumulants are in an
amazing agreement with our findings, see Table \ref{t2c} and Fig.
\ref{2c}.

 Notwithstanding the possibility that production of hadrons in {\ep}
annihilation may well be simpler than in hadron induced reactions, it may
nevertheless be instructive to relate our results to the measured
correlation data listed in Tables \ref{tw1c} and \ref{tw2c}.
 Inasmuch that the mean multiplicity increases with the number of
sources, the measurements consistent with our findings. 
 As well as in nucleus induced reactions, we found the decrease in the
two-particle rapidity correlations and the absence of three-pion rapidity
correlations.
 In two dimensions our results show a surviving of higher order
correlations and a change of the hierarchy not starting from two sources,
but later at larger number of them, the effects which one can see in the
data in nucleus-nucleus reactions. 
 All this demonstrates the effect of dilution of the correlations with
increased number of sources.

A quantitative comparison between our findings and the correlations
in nucleus-nucleus and hadron induced reactions is hard to make 
mainly because of the lack of information on the values 
of $K^{\rm av(1)}_q$ which should be derived from nucleon-nucleon
collisions data. In addition in many of the existing cumulant data
of the nucleus induced reactions have  
still too large errors.

 Recently the two-pion BEC have been studied \ct{bec-iss} in ${\bar {\rm
p}}$p reaction at centre of mass energy of 630 GeV as a function of
multiplicity by using the normalised cumulants method similar to the one
used here.
 In that analysis it has been found that the correlations 
of the cumulants of the like-sign pions as well as the opposite-sign pions 
decrease with the multiplicity $n_{ch}$. 
 From our analysis we expect the pair correlation to decrease as $1/S$,
where $S$ is the number of sources. This indicates that indeed the
multiplicity is at least partially proportional to $S$.
 The BEC dependence has also been investigated in the framework of the
totally coherent emission picture \ct{bec-coh} and in the
quantum optical approach \cite{bec-is} where the conclusions were that
these correlations are weaker as the multiplicity increases.

\section{SUMMARY AND CONCLUSIONS}
\la{sum}

 In the present work we investigated the effect of many emission sources
on the genuine correlations in mutihadron final state.
 To this end we adopted a procedure which should minimise the confusion
introduced by other variables like charged multiplicity. 
 For the genuine correlations measurement we utilised the normalised
cumulant method. 
 To simulate the situation of many sources event we did overlay Monte
Carlo generated hadronic Z$^0$ events treating them as one event. This \JT
\ 7.4 MC sample of some five million events, tuned to the OPAL data taken at
LEP1 on the Z$^o$ mass, has previously described rather well the measured
correlations in the {\ezh} data. 
 We studied the cumulants in one, rapidity, dimension and in two
dimensions of rapidity and azimuthal angle. 

The results shown here demonstrate that the cumulants, obtained from a
single-source events and from events of many sources, almost do not change
their basic structure with the decrease of the width of phase-space bins. 
 This means that the scaling is preserved although larger slopes are seen
to be in the case of one source compared with those for several sources. 

Due to source mixing the higher order cumulants are suppressed
both the 1-dimensional (rapidity) and the 2-dimensional
(rapidity $\times$ azimuthal angle) 
and diminish to zero as the number of sources increases. In both case
as $S$ increases the hierarchy is reversed and the cumulant of
the lower order $q = 2$ dominates. This happens 
in the one dimension case already when $S = 2$ whereas in the two
dimension occurs only at a higher $S$ values of around $S = 5$. 
 
The correlations observed are very well reproduced by assuming that
genuine correlations of the order $q$ can only exist when all the $q$
hadrons are emerging from the same source.
 Therefore the dilution of the genuine correlation signal is proportional
to the ratio of the probability that the $q$ hadrons will emerge from the
very same source. From simple combinatorial considerations this probability 
can then be  approximated by $1/S ^{q-1}$.
 Thus a measurement of the correlations of two sequential orders in 
$q$ can be used to estimate the average number of sources. 

The genuine correlations measured in hadron and nucleus induced reactions
do follow qualitatively the findings of our work.
 In particular in nucleus-nucleus reactions, where many sources are
expected to contribute to the final hadronic state, the rapidity cumulants
of the $q > 2$ orders are very small and indeed consistent with zero.
 The $q=2$ order of the 1-dimensional cumulants still survive but they 
also are getting smaller as the atomic number of the nuclei increases.
 In two dimensions, the cumulants are still different from zero 
even at the $q=4$ order, however their value decreases
and a change in the hierarchy takes place as the multiplicity
increases. The general observation that  
cumulants decrease as the multiplicity increases 
reassures the common notion that the higher the multiplicity the larger the
number of sources. 

 Our results may also be useful for the understanding of other types of
measured correlations like that obtained from the Bose-Einstein 
interferometry of two and more identical bosons.  
 It has been previously pointed out \ct{bec-w} that in the absence of
final state interactions the BEC of the e$^+$e$^- \to$
W$^+$W$^- \to hadrons$ will be half of that of the {\z} decay to hadrons. 
From our study it follows directly that the two-particle BEC, or any other
correlations, in the two-source reaction e$^+$e$^- \to$ W$^+$W$^- \to
hadrons$ should be reduced by a factor two as compared to that of
the hadrons emerging from one W-boson. 

\section*{\small\bf Acknowledgements}

We would like to thank the organisers of the Workshop on
Multiparticle Production for their very kind hospitality
and for the warm and pleasant atmosphere that they maintained 
over the whole conference duration.



{\small

\begin{thebibliography}{99}
\bi{revi} E.A. De Wolf,
           I.M. Dremin and W. Kittel, \jour{\PRp}{270}{96}{1}.
\bi{Ogc} {\OP} \col, G. Abbiendi {\ea}, \jour{\EPJ}{11}{99}{239}.
\bi{hAc} P. Carruthers, H. Eggers and I. Sarcevic, 
          \jour{\PRC}{44}{91}{1629}. 
\bi{na22} EHS/NA22 \col, N. Agababyan \ea, \jour{\ZP}{59}{93}{405};\\
          M. Charlet, Ph.D. Thesis, {\it Multiparticle Correlations in
$\pi^+$p and K$^+$p interactions at 250 GeV/$c$}, (Nijmegen Univ., 1994),
unpublished.
\bi{hhc} EHS/NA22 \col, N.M. Agababyan \ea, \jour{\PL}{332}{94}{458};\\
          E665 \col, M.R. Adams \ea, \jour{\PL}{335}{94}{535}. 
\bi{AAc2} P.L. Jain and G. Singh, \jour{\NPA}{596}{96}{700}. 
\bi{nfcd} EMU01 \col, M.I. Adamovich \ea, \jour{\NPB}{388}{92}{3}. 
\bi{isn} EMU01 \col, M.I. Adamovich \ea, \jour{\PRD}{47}{93}{3726}. 
\bi{AAc1} P.L. Jain, A. Mukhopadhyay, G. Singh, \jour{\ZP}{58}{93}{1}. 
\bi{is1} P. Lipa and B. Buschbeck, \jour{\PL}{223}{89}{465}.
\bi{is2} S. Barshay, \jour{\ZP}{47}{90}{199};\\
        D. Seibert, \jour{\PRD}{41}{90}{3381}.
\bi{ncf} EMU01 \col, M.I. Adamovich \ea, \jour{\PL}{263}{91}{539}, 
                                           \jour{\PL}{407}{97}{92};\\
         E802 \col, T. Abbott \ea,  \jour{\PRC}{52}{95}{2663}.   
\bi{hhcf} P. Carruthers, H. Eggers and I. Sarcevic, 
          \jour{\PL}{254}{91}{258}.
\bi{hhbec}  EHS/NA22 \col, N.M. Agababyan \ea, \jour{\ZP}{68}{95}{229}.
\bi{eebec} {\DE} \col, P. Abreu \ea, \jour{\PL}{355}{95}{415};\\ 
          {\OP} \col, K. Ackerstaff \ea, \jour{\EPJ}{7}{99}{379}.
\bi{AAbec} NA44 \col, H. B{\o}ggild \ea, \jour{\PL}{455}{99}{77}.
\bibitem{wacollab} WA98 \col, M.M. Aggarwal et al., hep-ex/0008018.
\bibitem{borghini} N. Borghini \ea, Phys. Rev C 62 (2000) 034902.
\bi{WWbec} See \eg, W. Kittel, talk given at the {\it QCD and
   Hadronic Interactions} session of the  34 Rencontres de Moriond 
   (Les Arcs, France, Mar. 1999), hep-ph/9905394 and references therein.
\bi{as_plb} See also G. Alexander and E.K.G. Sarkisyan, Phys. Lett. 
B 487 (2000) 215.  
\bi{bec-is} N. Suzuki and M. Biyajima, \jour{\PRC}{60}{99}{034903}.
\bi{bec-iss} B. Buschbeck, H.C. Eggers and P. Lipa, 
     \jourm{\PL}{481}{00}{187}.
\bi{qcd} I.M. Dremin, \jour{\UFN}{37}{94}{715};\\
         V.A.~Khoze and W.~Ochs, \jour{\IJ}{12}{97}{2949}. 
\bi{JT} T. Sj$\ddot{\rm o}$strand, \jour{\CP}{82}{94}{74}.
\bi{MCOd} J. Allison \ea, \jour{\NIM}{317}{92}{47}.
\bi{MCO} {\OP} \col, G. Alexander \ea, \jour{\ZP}{69}{96}{543}.
\bi{cum} P. Carruthers and I. Sarcevic, \jour{\PRL}{63}{89}{1562};\\
          E.A. De Wolf, \jour{\AP}{21}{90}{611}.
\bi{math} M.G. Kendall and A. Stuart, The Advanced Theory of Statistics
          (C. Griffin \& Co., London, 1969), Vol. 1;\\
          A.H. Mueller, \jour{\PRD}{4}{71}{150}.
\bi{bec-coh} M. Biyajima \ea, \jour{\PL}{386}{96}{297}.
\bi{bec-w} S.V. Chekanov, E.A. De Wolf, W. Kittel,
           \jour{\EPJ}{6}{99}{403}.

\end{thebibliography}
}
\end{document}